\documentclass[12pt]{iopart}
\usepackage{graphicx,psfrag}
\usepackage{cite,epsfig}

\newcommand{\pq}[1]{\left[{#1}\right]}
\newcommand{\pg}[1]{\left\{{#1}\right\}}

\newcommand{\EL}{\mathcal{L}}
\newcommand{\de}{{\mathrm d}}

\newcommand{\lij}{l_{ij}}
\newcommand{\E}{\mathrm{e}}
\newcommand{\fm}{f_\mathrm{max}}
\newcommand{\tm}{t_\mathrm{max}}
\renewcommand{\Re}{\mathrm{R}}
\renewcommand{\Im}{\mathrm{I}}
\newcommand{\I}{\mathrm{i}}

\begin{document}

\title{Work distribution in manipulated single biomolecules}

\author{A. Imparato}

\address{Department of Physics and Astronomy, University of Aarhus,
  Ny Munkegade, Building 1520, DK--8000 Aarhus C, Denmark}
\ead{imparato@phys.au.dk}
\author{L. Peliti}
\address{Kavli Institute for Theoretical Physics,
University of California,
Santa Barbara, CA~93106--4030\\
and\\
Dipartimento di Scienze Fisiche and Unit\`a INFN, 
Universit\`a ``Federico II'', Complesso Monte S.~Angelo, I--80126 Napoli (Italy)}
\ead{peliti@na.infn.it}
\pacs{82.37, 87.15, 05.70}

\begin{abstract}
We consider the relation between the microscopic and effective descriptions
of the unfolding experiment on a model polypeptide. We evaluate the probability
distribution function of the performed work by Monte Carlo simulations
and compare it with that obtained by evaluating the work distribution
generating function on an effective Brownian motion model tailored
to reproduce exactly the equilibrium properties. The agreement is satisfactory
for fast protocols, but deteriorates for slower ones, hinting at the existence
of processes on several time scales even in such a simple system.
\end{abstract}
\maketitle

\section{Introduction}
By means of atomic force microscopes and optical tweezers, several experimental groups have been able to control very precisely the force applied on proteins and nucleic acids.
The observation  of the unfolding behavior of these molecules under mechanical stress represents a powerful tool to recover structural properties of proteins and nucleic acids \cite{KSGB,rgo,cv1,DBBR,Ober1,Li_2000,exp_fc1,exp_fc2,NA_exp}.

Furthermore, in the case of small biomolecules, unfolding experiments represent an excellent test bed for a class of recently derived results, which go under the name of \emph{fluctuation relations}~\cite{JE,CRO,HumSza}.
These relations connect the energy exchanged by thermodynamical systems with their environment to their equilibrium properties, and
represent therefore an intriguing bridge between equilibrium and nonequilibrium.
The case of small biomolecules under external force is interesting in itself, since the fluctuations of the energy exchanged by these systems are of the order of their average thermal energy, and they are therefore an example of microscopic out-of-equilibrium systems, with very large thermal fluctuations, whose study is one of the current topical problem in statistical mechanics \cite{fel_rev}.

As a force is applied on a biomolecule, and it progressively unfolds, the external pulling device performs thermodynamical work on it.
By sampling this quantity over many repetitions of the unfolding experiments,
and taking advantage of suitable fluctuations relations \cite{JE,CRO,HumSza}, it has been possible to estimate experimentally
the free energy difference between the folded and the unfolded state of
a simple RNA hairpin~\cite{jarzexp} and the free energy landscape of some proteins as a function of the molecular elongation~\cite{HSK,ISV}.
Convergence in such estimate is dominated by the so-called outliers,
i.e., rare values of the work that are much smaller than the average.
The interest in the study of distribution functions  of work performed on biomolecules
during unzipping experiments and in particular of the distribution tails, 
is due to the need to estimate the frequency of the rare events that ensure validity of the fluctuation relations. 

The evaluation of the work probability distribution function (PDF) for a manipulated system requires in principle
the solution of an evolution equation of complexity equivalent to the equation
for the microscopic dynamics. Since this equation involves a large number of
degrees of freedom even for comparatively small systems like a polypeptide, it
is customary to describe its dynamics by a small set of collective coordinates
undergoing a Brownian diffusion process ~\cite{HumSza,noi2,noi3,sei, HumSza1}. 
It is worth to note that in their works, Hummer \textit{et al.}~\cite{HumSza1} showed that  if the molecular unfolding is described as a  one dimensional diffusion process along a structured energy potential, the force-induced rupture rate exhibits a behaviour which is much more complex than the widely used approach based on the Bell's formula.
Here we investigate the relation
between the two levels of description on a simple model of a polypeptide
unfolding experiment, and differently from \cite{HumSza1}, we focus on the 
description of the work distribution, rather than on the description of the unfolding rate.

\section{The work distribution}
\label{uno}
Let us consider a system whose microscopic state is identified by the 
variable $x$, where $x$ can also indicate a collection of microscopic coordinates, e.g., the positions and momenta of the particles which make up the system.
We shall assume that the system evolves according to a general dynamic process, parametrized by a parameter $\mu$, which can be
manipulated according to a fixed protocol $\mu(t)$. The evolution can be deterministic or stochastic, but we shall assume that,
for any given value of $\mu$, there is a well-defined equilibrium distribution that can be represented in the Boltzmann-Gibbs form
\begin{equation}
     P^\mathrm{eq}_\mu(x)=\frac{\E^{-\beta H(x,\mu)}}{Z_\mu},
\end{equation}
a relation which \textit{defines} the Hamiltonian $H(x,\mu)$ and the partition function $Z_\mu=\int \de x\;\E^{-\beta H(x,\mu)}$. Here $\beta=1/k_BT$, where $T$ is the absolute
temperature and $k_B$ Boltzmann's constant.
Thus $H(x,\mu)$ depends explicitly on the time only via $\mu(t)$.
The probability distribution function (PDF) $P(x,t)$ of the microscopic state $x$ evolves according to the Liouville-like partial differential equation
\begin{equation}
\partial_t P(x,t)=\EL_\mu\, p(x,t),
\label{eq_PDF}
\end{equation} 
where $\EL_\mu$ is a linear differential operator, compatible
with the equilibrium distribution of the system for any fixed value of $\mu$:
$\EL_\mu P^\mathrm{eq}_\mu=0$, $\forall\mu$.

The external manipulation of the system via $\mu$ leads to an energy exchange
with the environment. According with the usual conventions in statistical 
mechanics (see, e.g.,~\cite{Luca}), the fluctuating work $W$ performed on
the system, given the manipulation protocol $\mu(t)$ and the microscopic
trajectory $x(t)$, is given by
\begin{equation}
     W=\int_0^t\de t'\; \dot\mu(t')\,\partial_\mu H(x,\mu)|_{x(t'),\mu(t')}.
\end{equation}
Under these hypotheses, the time evolution of the joint PDF $\phi(x,W,t)$
of the microscopic state $x$ and the total work $W$ performed on the system
is governed by the partial differential equation \cite{noi2,noi3}
\begin{equation}
\partial_t \phi(x,W,t)=\EL_\mu \phi(x,W,t)- \dot\mu(t)\,\partial_\mu H(x,\mu)|_{\mu(t)}\, \partial_W \phi(x,W,t).
\label{eq_PDFw}
\end{equation} 
In order to simplify the analysis, one evaluates the 
generating function $\psi(x,\lambda,t)$ of the distribution of $W$, defined by
\begin{equation}
\psi(x,\lambda,t)=\int \de W\, \E^{\I\lambda W} \phi(x,W,t),
\label{eq_gen}
\end{equation} 
so that eq.~(\ref{eq_PDFw}) becomes
\begin{equation}
\partial_t \psi(x,\lambda,t)=\EL_\mu \psi(x,\lambda,t)+\I\lambda\, \partial_t H \, \psi(x,\lambda,t).
\label{eq_psi}
\end{equation} 
This equation can be solved explicitly if the system is characterized by discrete states: in ref.~\cite{noi1}, e.g., an RNA hairpin was modelled as a three-state system
and the  PDF of the work, done on the molecule by an external mechanical force, was evaluated numerically.

One can evaluate the solution of eqs.~(\ref{eq_gen}--\ref{eq_psi}) for real $\lambda$, starting from the initial condition $\psi(x,\lambda,0)=P^{\mathrm{eq}}(0)$,
$\forall\lambda$. Thus, since $\phi(x,W,t)$ is real, we have 
$\psi(x,-\lambda,t)=\psi^*(x,\lambda,t)$, and we can restrict ourselves to the half-line $\lambda \ge 0$. Then one can separate eqs.~(\ref{eq_psi}) 
into  two equations, one for the real part $\psi_\Re$  and one for the imaginary part $\psi_\Im$ of $\psi$, obtaining
\begin{eqnarray}
\partial_t \psi_\Re(x,\lambda,t)=\EL_\mu \psi_\Re(x,\lambda,t)-\lambda\, \partial_t H \, \psi_\Im(x,\lambda,t), \label{psir}\\
\partial_t \psi_\Im(x,\lambda,t)=\EL_\mu \psi_\Im(x,\lambda,t)+\lambda\, \partial_t H \, \psi_\Re(x,\lambda,t)\label{psii}.
\end{eqnarray} 

Once the function $\psi(x,\lambda,t)$ has been obtained, the joint PDF $\phi(x,W,t)$
is given by the inverse Fourier transform of $\psi(x,\lambda,t)$.
The unconstrained work PDF can be then evaluated from the relation $\Phi(W,t)=\int \de x\, \phi(x,W,t)$.

An important special case obtains when one considers a system described by few degrees of freedom, which is in contact
with a large heat reservoir.
In this case, it is often warranted to assume that the microscopic state $x$ performs a Brownian motion~\cite{HumSza,noi2,noi3,sei, HumSza1}, 
and thus the operator $\EL_\mu$ has the form of a Fokker-Planck (FP) differential operator:
\begin{equation}
\EL_\mu\cdot{}= \Gamma \frac{\partial}{\partial x} \pq{\partial_x H(x,\mu)\cdot{} + k_B T \frac{\partial}{\partial x}\cdot{}},
\label{fpop}
\end{equation}  
where we take into account the Einstein relation between the diffusion and
the kinetic (mobility) coefficients $D=\Gamma k_B T$.

\section{Collective coordinates}
\label{due}
In an unfolding experiment, where a mechanical force is applied to one or both  the free ends of a biopolymer, the work can be sampled by monitoring the extension of the molecules at different times \cite{jarzexp,HSK,ISV}. In this situation we face the following problem.
Equations (\ref{eq_PDF}), (\ref{eq_PDFw}) and (\ref{eq_psi}) describe the dynamics  of a system at a very detailed, microscopic level. On the other hand, the behavior of the system is accessed only via the measurement of a few, and most often only one, observables, such as the elongation. Moreover, the microscopic ``Liouville'' operator is not generally known with sufficient confidence. In any case, the explicit solution of the evolution equations becomes unfeasible as soon as more than a few degrees of freedom have to be considered.

Thus one considers descriptions of the system through some  experimentally accessible collective coordinates. In the case of biopolymers, one typically chooses the end-to-end length $L$. Its equilibrium distribution
is determined by the effective free energy, defined by
\begin{equation}
F(L,\mu)=-\beta^{-1}\ln\int \de x\; \delta(L(x)-L)\, \E^{-\beta H(x,\mu)}.
\label{fel}
\end{equation} 
which plays the role of an effective hamiltonian.
The dependence of this free energy on $L$ is the target of several experimental studies, performed by using a suitable fluctuation relation \cite{HSK, ISV}.

It should be possible in principle to obtain the evolution for the collective coordinate PDF, and thus the work distribution,
by projecting the microscopic ``Liouville'' equations on the space spanned by the collective coordinates~\cite{Mori}.
However, one would then in general obtain complicated non-Markovian evolution equations, whose parameters will depend
on unknown details of the underlying microscopic dynamics.
In other words, even if eqs.~(\ref{eq_PDF}--\ref{eq_psi}) were exact, it would be hard to derive the explicit
equations governing the time evolution of the PDF $P(L,t)$ and the joint PDF $\phi(L,W,t)$.

Thus, in the present work, we make the following ansatz: we assume that the coordinate $L$ performs a Brownian motion in an effective potential, which is given by the free energy landscape (\ref{fel}).
This implies that   the evolution operator $\EL_\mu$ for the coordinate $L$ is a FP operator of the form of eq.~(\ref{fpop}), 
where $x$ has to be replaced by $L$, and $H(x,\mu)$ has to be replaced by $F(L,\mu)$, as given by eq.~(\ref{fel}). 
This is a bold summary of the underlying microscopic process. In the resulting model, the form of the evolution equation is
constrained, but the value of the kinetic coefficient $\Gamma$ is still free. We shall take advantage of this degree of freedom
and check whether it allows us to describe the behavior of the work PDF with sufficient fidelity.

\section{Lattice model of proteins}
\label{sec_mod}
\begin{figure}[ht]
\center
\psfrag{lij}[tt][tt][1.2]{$\lij$}
\psfrag{L}[tt][tt][1.2]{$L$}
\psfrag{f}[cb][cb][1.2]{$f$}
\psfrag{zeta}[lt][lt][1.2]{$\zeta$}
\includegraphics[width=8cm]{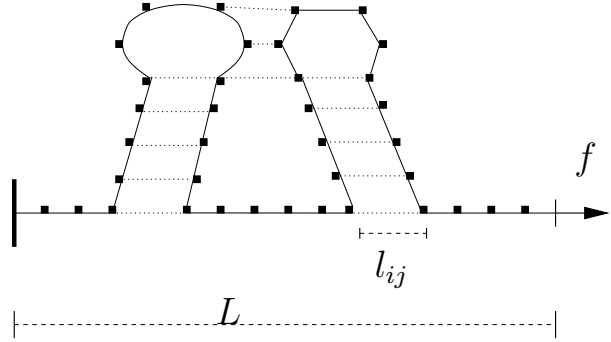}
\caption{Cartoon of the model protein, with a force applied to
one of the free ends. Dots denote amino~acids and dashed lines
denote contacts.}
\label{molec}
\end{figure}

In the present section we consider a lattice model for proteins under mechanical load, that, in spite of its simplicity, is able
to reproduce in great detail the outcome of experiments performed on real proteins~\cite{IPZ07,IPZ_a07,IP08}.
In this model, the state of a $N+1$ aminoacid protein is defined by the set of discrete variables $\pg{m_k}$, $k=1\dots N$. 
These binary variables 
take the value $m_k=0$ ($m_k= 1$), if the peptide bond is in the non-native (native) configuration.
Then the effective hamiltonian reads
\begin{equation}
 H_\mathrm{eff}(\{m_k\},L,f)=\sum_{i=1}^{N-1} \sum_{j=i+1}^N
\epsilon_{ij} \Delta_{ij} \prod_{k=i}^j m_k -f\cdot L(\{m_k\},\{\sigma_{ij}\}),
\label{heff}
\end{equation} 
where
\begin{equation}\label{length:eq}
     L(\{m_k\},\{\sigma_{ij}\})=\sum_{0\le i<j\le N+1}l_{ij}\sigma_{ij}S_{ij}(m)
\end{equation}
is the end-to-end distance in the configuration $x=(\{m_k\},\{\sigma_{ij}\})$,
projected on the direction of the applied force, as shown in fig.~\ref{molec}.
In this hamiltonian the quantity $\epsilon_{ij}\le0$ represents
the interaction energy between the residues $i$ and $j+1$,
$l_{ij}$ is the length of the native strand of peptide bonds between residues
$i$ and $j$, or the length of the single
non-native bond $i,i+1$, and the binary variable $\sigma_{ij}$ is equal to 1
if the strand is parallel, and to $-1$ if it is antiparallel to the
applied force. The quantity $S_{ij}(m)$ is equal to 1 if the  polypeptide
strand starting at $i$ and ending at $j$ is all in the native state, and
is flanked by bonds in the non-native state, and vanishes otherwise, 
as explained in \cite{IPZ07,IPZ_a07}. 
For a given protein, the parameters $\epsilon_{ij}$ and $\lij$ 
are obtained by analyzing the protein native structure, as given in the Protein Data Bank (PDB).

Here, we consider in particular the polypeptide PIN1 (PDB code 1I6C) which is made up of 39 aminoacids, 
at a reduced temperature $\tilde T=6$ (cf.\ ref.~\cite{IPZ_a07} for a detailed discussion on the temperature and force scales).

The unfolding experiments are simulated using the Monte-Carlo Metropolis algorithm with the hamiltonian (\ref{heff}), 
where the  external force varies linearly with time $f=r\cdot t$.
For each unfolding trajectory, the system is prepared in equilibrium with vanishing force, and then at $t=0$, the 
force starts increasing with rate $r$. For practical purposes, we define the force rate as $r=\fm/\tm$, keeping constant $\fm=10$, and varying $\tm$. We have simulated 10000 unfoldings for each value of the rate $r$. We have then sampled the work $W=\int_0^{\tm} \partial_t H_\mathrm{eff}(\{m_k\},\{\sigma_{ij}\},f(t))$ performed on the molecule and evaluated the work histograms.

For the model protein here considered, the free energy landscape $F_0(L)$, as defined by
\begin{equation}
F_0(L)=-k_B T \ln \pg{\sum_x \exp\pq{-\beta H_\mathrm{eff}(x,f=0)} \delta (L(x)-L)},
\label{ef0}
\end{equation} 
can be exactly calculated \cite{IPZ07,IPZ_a07}, and  the time-dependent
landscape reads therefore $F(L,f(t))=F_0(L)-f(t)L$. 
In eq.~(\ref{ef0}), $x$ represents the microscopic state of the model, 
i.e., the collection of the variables $\{m_k\}$ representing the state of the bonds, and of the variables $\{\sigma_{ij}\}$ representing the orientation of
the strands with respect to the reference direction.

\begin{figure}[ht]
\center
\psfrag{f0}[lt][lt][.8]{$f=0$}
\psfrag{f2}[lt][lt][.8]{$f=2$}
\psfrag{f3}[lt][lt][.8]{$f=3$}
\psfrag{L}[ct][ct][1.]{$L$ (nm)}
\psfrag{F}[ct][ct][1.]{$F\,  (k_B T)$}
\includegraphics[width=8cm]{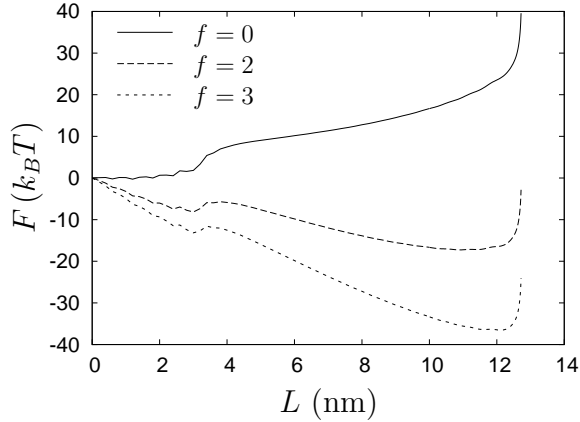}
\caption{Free energy landscape $F(L,f)$ for the model PIN1 polypeptide, for different value of the external force. The force is expressed in reduced units, see ref.~\cite{IPZ_a07}.}
\label{land}
\end{figure}

The free energy landscape $F(L,f)$ of this polypeptide in plotted in fig.~(\ref{land}), for different values of the external force.
Inspection of this figure indicates that at vanishing external force, the potential is almost flat for $L\le 3$~nm, while for larger values of the force a minimum appears at $L^*\simeq 12.6$~nm, whose position is practically independent of $f$, 
indicating that $L^*$ represents the length of the fully stretched molecule \cite{IPZ_a07}.
As discussed in section~\ref{due}, we will take this energy landscape as effective potential in the differential operator 
(\ref{fpop}).

\begin{figure}[ht]
\center
\psfrag{lam}[ct][ct][1.]{$\lambda$}
\psfrag{pr}[rc][rc][1.]{$\Psi_\Re$}
\psfrag{pi}[rc][rc][1.]{$\Psi_\Im$}
\includegraphics[width=8cm]{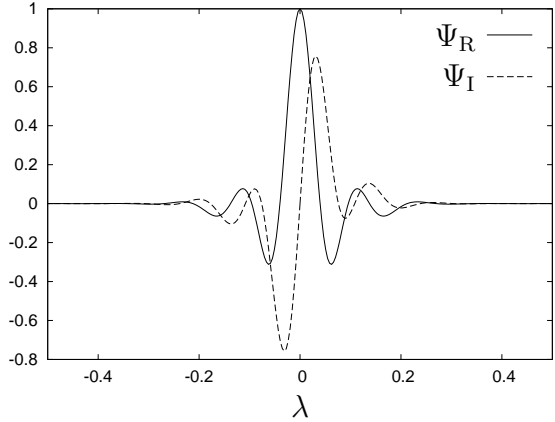}
\caption{Real and imaginary parts
 of the unconstrained generating function $\Psi(\lambda,\tm)=\Psi_\Re(\lambda,\tm)+\mathrm{i}\Psi_\Im(\lambda,\tm)$ vs.\ $\lambda$, as obtained from the numerical solutions of eqs.~(\ref{psir},\ref{psii}), with  $\Gamma=13.75$, and $r=1$ ($\tm=10$). }
\label{figpsi}
\end{figure}

\begin{figure}[ht]
\center
\psfrag{la}[rc][rc][.8]{(a)}
\psfrag{lb}[rc][rc][.8]{(b)}
\psfrag{lc}[rc][rc][.8]{(c)}
\psfrag{ld}[rc][rc][.8]{(d)}
\psfrag{W}[t][cc][1.]{$W\, (k_B T)$}
\includegraphics[width=6cm]{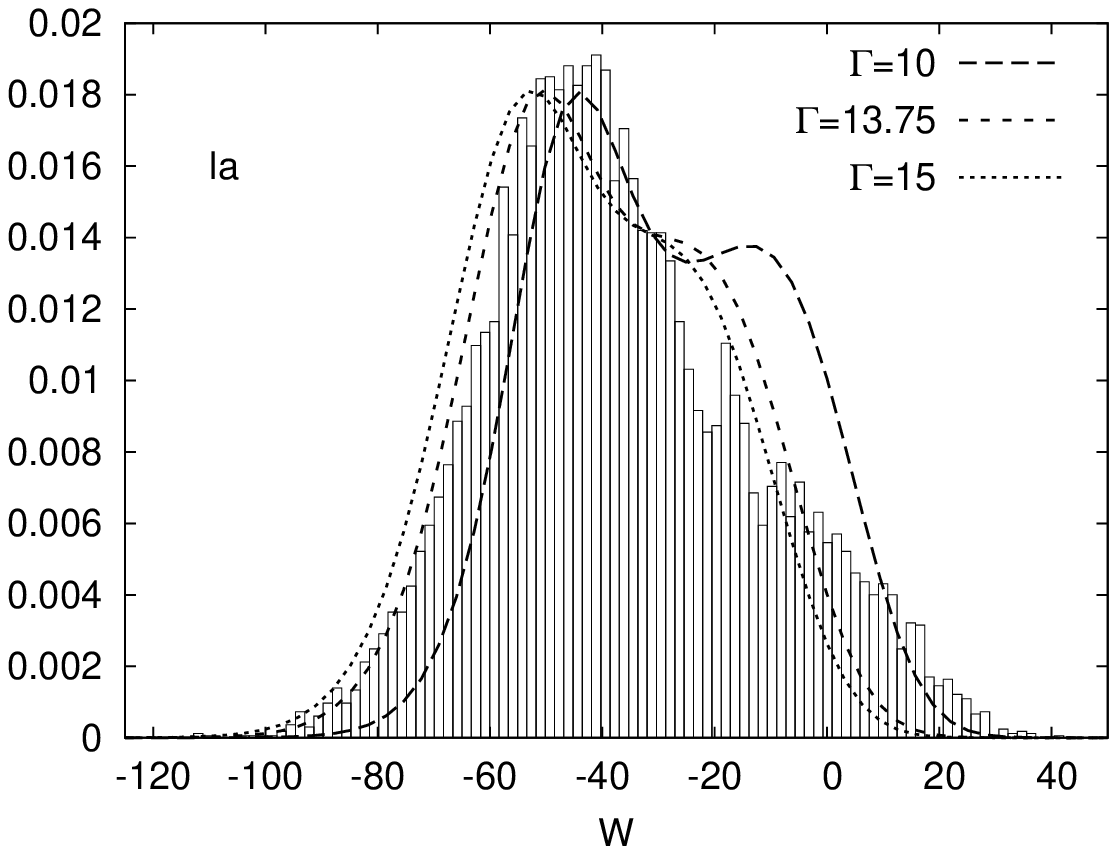}
\includegraphics[width=6cm]{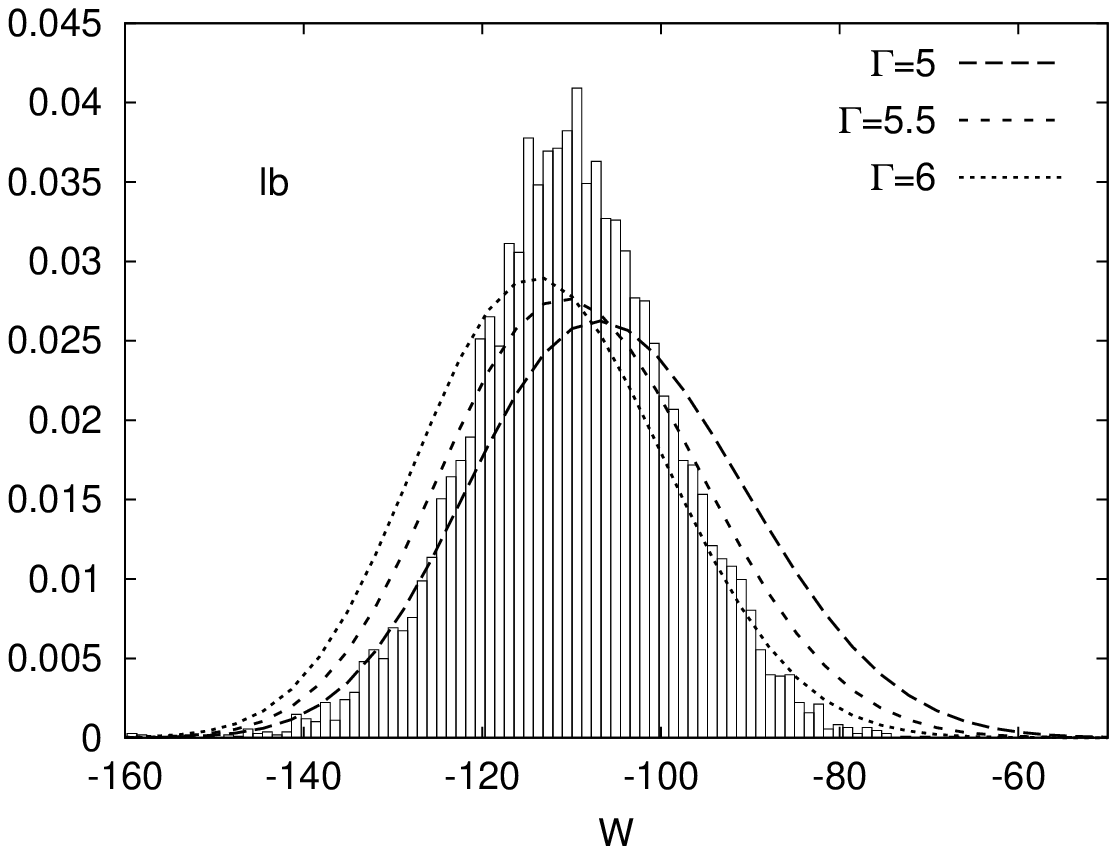}\\[0.2cm]
\includegraphics[width=6cm]{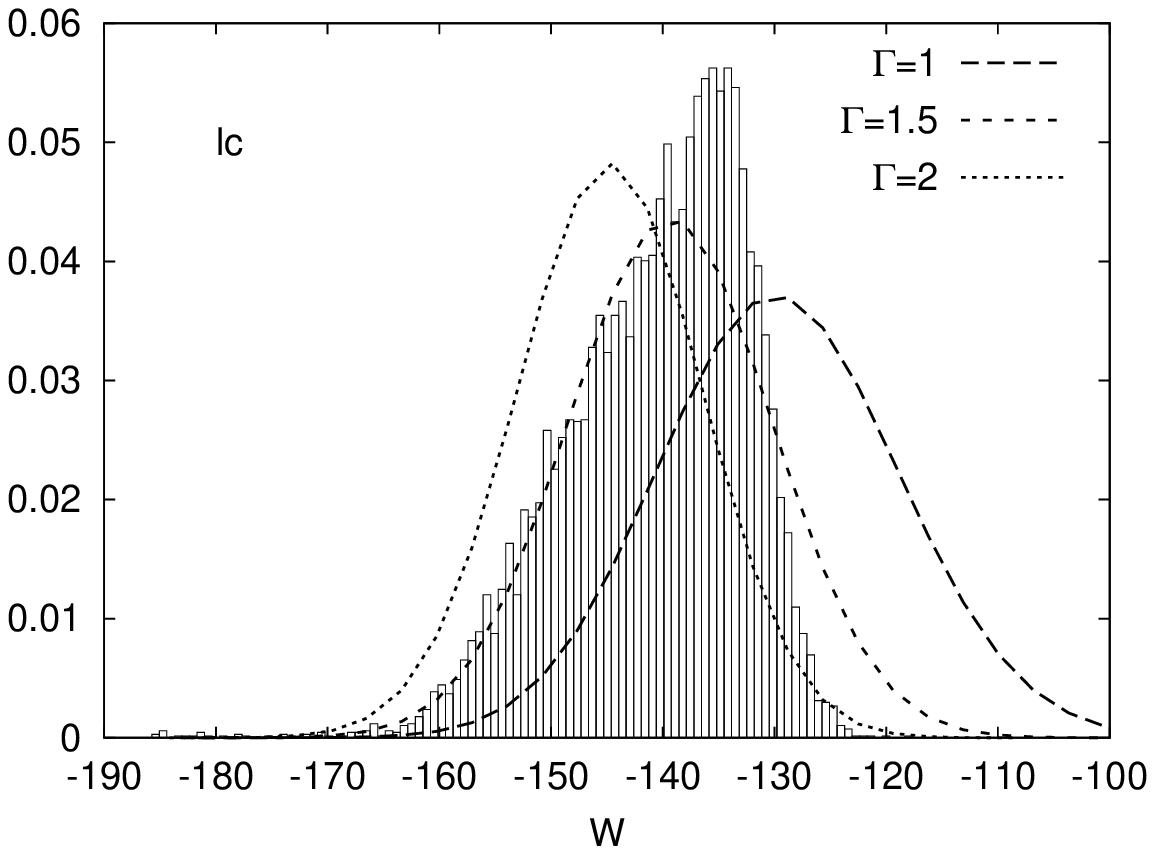}
\includegraphics[width=6cm]{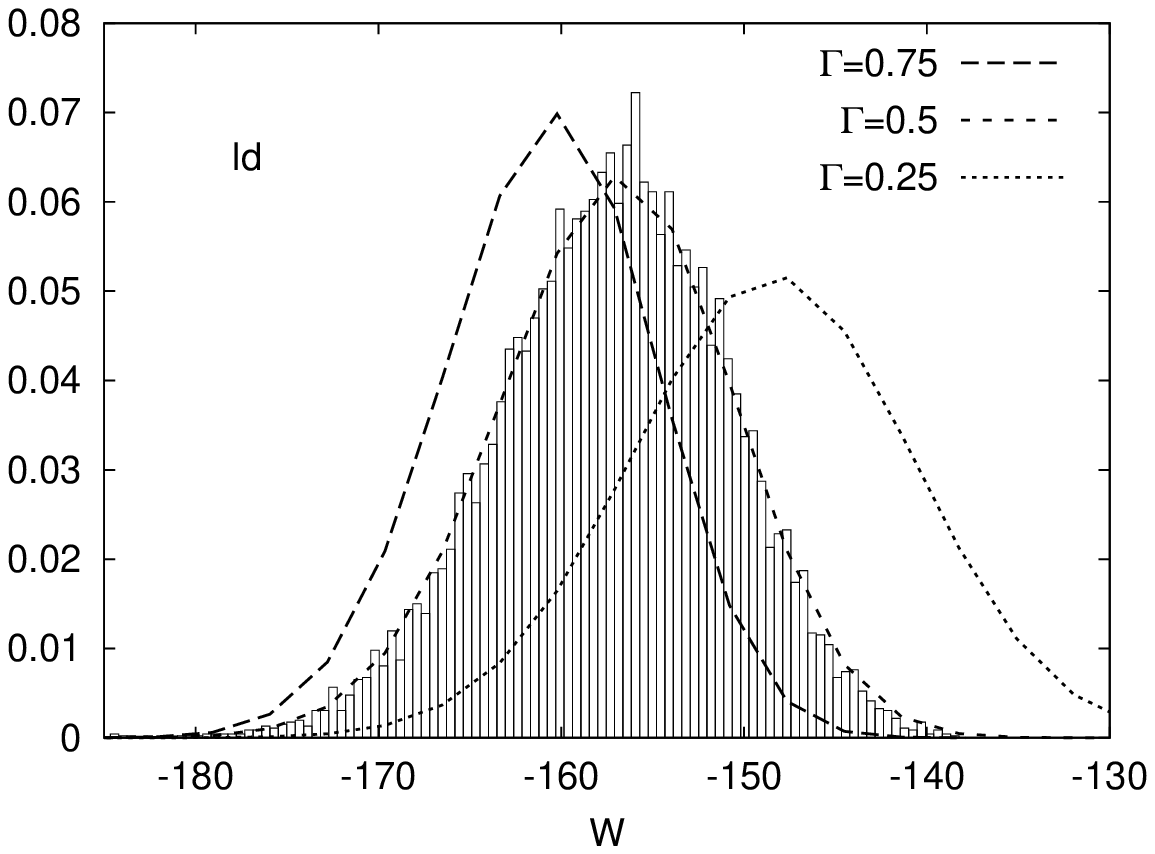}
\caption{Work PDF for the model protein discussed in the text, for different values of the force rate $r$: $r=1$ (a), $r=0.1$ (b), $r=0.01$ (c), $r=0.001$ (d).
The value of $\Gamma$ shown in the legend of each figure, corresponds to the value used to solve numerically eqs.~(\ref{eq_psi}).}
\label{wPDF}
\end{figure}

In figure~\ref{wPDF}
we show the histograms of the work PDF, as obtained by the simulations discussed above, for four different values of the manipulation rate $r=\fm/\tm$.
In the same figure, we plot the probability distribution function as obtained 
by solving a discretized version of the equations (\ref{eq_PDFw}--\ref{eq_psi}).

In this approach the equations take the form of a master
equation, in which the states are identified by an integer $i$, where $L_i$
is the polymer length measured in units $\Delta L=L_\mathrm{max}/N$, where 
$L_\mathrm{max}$ is the maximum length that can be obtained in the 
lattice model, and $N=126$. Positive and negative values of $L$ are considered.
The transition rates $W_{i\longrightarrow i\pm 1}$ are defined to match
those of a Metropolis process with an attempt frequency equal to $\Gamma$:
\begin{equation}\fl\qquad
     W_{i\longrightarrow j({=} i\pm 1 )}(t)=\Gamma\times\cases{1,&if $F(L_j,f(t))\le F(L_i,f(t))$;\cr
\E^{-\beta (F(L_j,f(t))-F(L_i,f(t)))},&otherwise.}
\end{equation}
The resulting equations can then be solved, by a classic Runge-Kutta method, when a definite value is assigned to the kinetic parameter $\Gamma$.
One then evaluates the unconstrained generating functions $\Psi_{\Re,\Im}(\lambda,\tm)=\int \de L\, \psi_{\Re,\Im}(L,\lambda,\tm)$. These functions are plotted
in fig.~\ref{figpsi}, for $\Gamma=13.75$, and $r=1$ ($\tm=10$).
As discussed in section~\ref{uno}, the unconstrained work PDF $\Phi(W,\tm)$ is finally obtained by inverting eq.~(\ref{eq_gen}).

For each value of $r$ we consider different values of the kinetic coefficient  $\Gamma$, and choose  the value which most closely reproduces the simulated
histograms. Indeed, even if the microscopic process goes on with a well-defined
characteristic attempt frequency, this is not the case for the effective 
process described by the FP equation. In order for $x$ to change, the 
microscopic variables $\{m_k,\sigma_{ij}\}$ must change. Their rate of change
will depend on an Arrhenius factor depending on the actual energy difference
due to the change of the particular value one is looking at. This factor will depend on the instantaneous value of the applied force, as well as on the 
overall state of the chain, and will not be a function only of $L$. 
We find nevertheless that it is possible to identify a value of $\Gamma$ which yields a reasonably good agreement for the faster protocols ($r=1$).
This value decreases as the manipulation speed decreases, showing
that in the slower manipulations there is a larger frequency of microscopic
processes that does not show up in changes of $L$.
For slower protocols ($r=0.1,\,0.01$), while one can match the mean value of the calculated work PDF that obtained with simulations, the shape
of the calculated distribution differs from the histograms.
Apparently
the intrinsic rate of the microscopic processes in these protocols cannot
be represented by a single attempt frequency, while it can for the faster protocols, which are dominated by simple ``snatching off'' of native regions. For the slowest manipulations ($r=0.001$) the work distribution becomes Gaussian. In this case, the Jarzynski identity \cite{JE,CRO} implies that its average $W_0$, its variance $\sigma^2_W$ and
the free-energy change $\Delta F$ must be related by
\begin{equation}
  W_0=\Delta F + \frac{\beta \sigma^2_W}{2}.
\end{equation}
Thus it is be possible to recover the distribution by fitting the single
parameter $\Gamma$, as we can see from the curves for $r=0.001$.

\begin{figure}[h]
\center
\psfrag{r}[ct][ct][1.]{$r$}
\psfrag{D}[ct][ct][1.]{$D$}
\includegraphics[width=8cm]{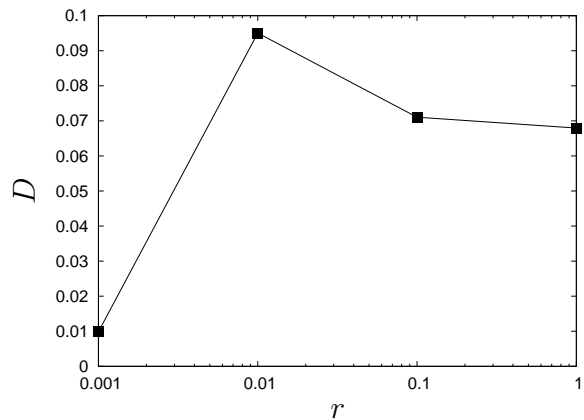}
\caption{Kolmogorov--Smirnov distance $D$ between the work distributions as obtained by simulations and by eqs.~(\ref{eq_PDFw}--\ref{eq_psi}), as a function of the loading rate $r$.}
\label{ksfig}
\end{figure}

In order to compare quantitatively the work PDFs as obtained by simulations and by solution of eqs.~(\ref{eq_PDFw}--\ref{eq_psi}), for each value of $r$ we exploit the Kolmogorov--Smirnov test~\cite{KS}. Thus, one evaluates the maximal
distance $D$ between the cumulative distributions of the two work PDFs:
\begin{equation}
D=\mathrm{sup}_x |\chi^{\mathrm{exp}}(W)-\chi^{\mathrm{theo}}{(W)}|,
\end{equation} 
where $\chi^{\mathrm{\alpha}}(W)=\int_{-\infty}^{W} \de  W\, \Phi^{\mathrm{\alpha}}(W)$, $\alpha=$exp/theo,
and where $\Phi^{\mathrm{exp}}(W)$ is the histogram as obtained by simulations, and
$\Phi^{\mathrm{theo}}(W)$ is the expected distribution, obtained with the procedure described in section~\ref{uno}.
The quantity $D$ is plotted in fig.~\ref{ksfig} as a function of $r$. Inspection
of this figure suggests that the smallest values of $D$ are obtained for $r=1$ and  $r=0.0001$, as indicated by a qualitative analysis of fig.~\ref{wPDF}.

\section{Discussion}
In this work we have investigated on a simple example the relation between the
work PDF obtained for the same system via its microscopic dynamics and
an effective Brownian dynamics. We found that the
Brownian dynamics works reasonably well for the faster protocols, but is off the
mark for slower ones, hinting at the existence of several dynamical time scales
in the relaxation of a moderately complex manipulated system. 
Thus, particular care has to be taken when comparing experimental outcomes with the results of numerical simulations, when the unfolding of a biopolymer is modelled as a biased one-dimensional Brownian process.

\ack
LP completed this research while visiting the Kavli Institute for Theoretical Physics within the program \textsl{FLUCTUATE08}.
This research was supported in part by the National Science Foundation under Grant No.~PHY05-51164. 
AI is grateful to A.~Pelizzola and M.~Zamparo for introducing him to the lattice model discussed in section~\ref{sec_mod}.

\end{document}